\documentclass[english,prl,twocolumn]{revtex4-1}
\usepackage[T1]{fontenc}
\usepackage[latin9]{inputenc}
\usepackage{color}
\usepackage{amsmath}
\usepackage{graphicx}



\usepackage{babel}
\begin{document}
\title{Steady thermodynamic fundamental relation for the interacting system in a heat flow}
\author{Robert Ho\l yst}
\email{equal contribution; rholyst@ichf.edu.pl}

\author{Karol Makuch}
\email{equal contribution; kmakuch@ichf.edu.pl}

\affiliation{Institute of Physical Chemistry, Polish Academy of Sciences Kasprzaka
44/52, 01-224 Warszawa}
\author{Konrad Gi\.{z}y\'{n}ski}
\affiliation{Institute of Physical Chemistry, Polish Academy of Sciences Kasprzaka
44/52, 01-224 Warszawa}
\author{Anna Macio\l ek}
\affiliation{Institute of Physical Chemistry, Polish Academy of Sciences Kasprzaka
44/52, 01-224 Warszawa}
\affiliation{Max-Planck-Institut f{\"u}r Intelligente Systeme Stuttgart, Heisenbergstr.~3,
D-70569 Stuttgart, Germany}
\author{Pawe\l{} J. \.{Z}uk}
\affiliation{Institute of Physical Chemistry, Polish Academy of Sciences Kasprzaka
44/52, 01-224 Warszawa}
\affiliation{Department of Physics, Lancaster University, Lancaster LA1 4YB, United
Kingdom}
\begin{abstract}

There is a long-standing question of whether it is possible to extend the formalism of equilibrium thermodynamics to the case of non-equilibrium systems in steady states. We have made such an extension for an ideal gas in a heat flow [Ho\l{}yst \emph{et al.}, J. Chem. Phys. 157, 194108 (2022)]. 
Here we investigate whether such a description exists for the system with interactions: the Van der Waals gas
in a heat flow. 
We introduce the parameters of state, each associated with a single way of changing energy.
The first law of non-equilibrium thermodynamics follows from these parameters. 
The internal energy $U$ for the non-equilibrium states has the same form as in equilibrium thermodynamics.
For the Van der Waals gas, $U(S^*, V, N, a^*,b^* )$ is a function of only 5 parameters of state (irrespective of the number of parameters characterizing the boundary conditions): 
the entropy $S^*$, volume $V$, number of particles $N$, and the rescaled Van der Waals parameters $a^*$, $b^*$. 
The state parameters, $a^*$, $b^*$, together with $S^*$, determine the net heat exchange with the environment.
 
\end{abstract}
\maketitle

\section{Introduction}

Determination of energy and its changes induced by heat or work are necessary to understand systems such as combustion engines or the earth's atmosphere with weather phenomena. 
When an equilibrium state approximates a system state, thermodynamics allows 
one to predict the system's behaviour by using energy as a function of a few parameters of state and a few principles.
In particular, the first law of thermodynamics
\citep{Thermodynamics_and_an_Introduction_to_Thermostatistics_2ed_H_Callen}  represents a global energy conservation law. 
The energy, $U(S, V, N)$  is a function of entropy, $S$, volume, $V$, 
and the number of molecules, $N$. Each variable is related 
to one independent way of energy exchange: heat, work, and change in the amount of matter.

However, a similarly simple theory does not exist for non-equilibrium
systems in steady (stationary) states. There is no description similar
to thermodynamics that grasps the energy transfer to the system in
terms of a few global parameters. 
One of the most straightforward
non-equilibrium cases is a steady heat flow. 
The appearance of the
heat flow opens many research directions belonging to various fields
of physics. 
Rational and extended thermodynamics focus on local transport equations \citep{jou2020relationships}. Irreversible thermodynamics
formulates thermo-hydrodynamic descriptions with local equations of
state and mass, momentum, and energy balance \citep{Groot_Mazur_Non-equilibrium_thermodynamics}.
Sometimes it is possible to represent governing equations in terms
of variational principles \citep{woo2002variational,Introduction_to_thermodynamics_of_irreversible_processes_Ilya_Prigogine,Non-equilibrium_Thermodynamics_Field_Theory_and_Variational_Principles_by_Dr_Istvan_Gyarmati,onsager1931reciprocal},
which determine the profile of thermodynamic fields (such as temperature).

The issue closely related to the studies mentioned above is whether we can represent the energy of the non-equilibrium system as a function of a few global parameters.
The answer to this question would lead to a description similar to classical equilibrium thermodynamics. 
The existence  of such a thermodynamic-like description for steady-state systems has been considered in various studies 
\citep{Introduction_to_thermodynamics_of_irreversible_processes_Ilya_Prigogine,landauer1978d,maes2019nonequilibrium,daivis2003steady,daivis2008thermodynamic,daivis2012thermodynamic}.
The progress \citep{komatsu2008steady,komatsu2011entropy,maes2014nonequilibrium,nakagawa2017liquid}
in this field is limited to small temperature differences and low heat fluxes.
The recent papers on this topic carry the conviction that general rules exist in non-equilibrium thermodynamics.
But scepticism regarding the usefulness of the equilibrium-based entropy \citep{lieb2013entropy} or even the existence of a description in terms of thermodynamic-like potentials \citep{jona2014thermodynamics}
also appears.

Lieb and Yngwasson \citep{lieb2013entropy} expressed scepticism regarding the use of entropy by suggesting heat as a primary quantity. 
It requires a generalization of heat for steady states. 
But how can it be generalized, e.g., for
a steady gas between two plates with heat flow in a perpendicular direction? Thermo-hydrodynamic equations describe the system, so the heat flowing through the surface is well-defined. 
This applies both for a steady state
and when the system passes from one stationary state to another. 
In a steady state, the same amount of heat enters through one plate
and leaves on the opposite side.
The net heat vanishes. 
But the net heat may flow to the system during the transition between steady states.
This reasoning leads to a concept of heat measured in transition between
stationary (steady) states. 
It is a particular case of the excess heat discussed by Oono and Paniconi \citep{oono1998steady}. 
In 2019 Nakagawa and Sasa \citep{nakagawa2019global} noticed that the excess heat concept defined by Oono and Paniconi had yet to be further utilized by other researchers. 
We adopt the term net (or excess) heat to name the heat that enters the system and changes its internal energy during the transition between steady states.
We note that in literature, the excess heat has other meanings \citep{chiba2016numerical}.

Our recent investigations of an ideal gas in a steady state with a heat flow showed a surprising result \citep{holyst2022thermodynamics}.
We proved that the net heat has an integrating factor and rigorously calculated non-equilibrium 'entropy` and non-equilibrium temperature. 
This entropy determines steady adiabatic insulation during transitions between stationary states.
However, it is not clear whether the non-equilibrium entropy exists beyond the ideal gas approximation. 
We continue research to formulate global steady thermodynamics using Van der Waals gas as an example of an interacting system. 
First, from the thermo-hydrodynamic
equations, we derive the global energy balance. 
Next, we show that it is possible to represent the non-homogeneous Van der Waals gas in a heat flow with equations formally identical to the equations of state for the Van der Waals gas in equilibrium. 
This procedure (named mapping) defines the parameters of the state for the non-equilibrium system in the steady state. 
We also show that the net heat does not have an integrating factor as proposed by Oono and Paniconi \citep{oono1998steady}.
Instead, the net heat is represented by two independent thermodynamic parameters of state in the Van der Waals gas.

\section{Van der Waals gas in equilibrium}

We consider the Van der Waals fluid described by the following fundamental thermodynamic relation \citep{Thermodynamics_and_an_Introduction_to_Thermostatistics_2ed_H_Callen}
\begin{equation}
	U=N\left(\frac{V}{N}-b\right)^{-\frac{1}{c}}\exp\left[\frac{S-Ns_{0}}{cNk_{B}}\right]-a\frac{N^{2}}{V}.
	\label{eq:fundamental eq}
\end{equation}
It binds together thermodynamic state functions, i.e., energy $U$, entropy $S$, volume $V$, and a number of particles $N$, with two interaction parameters $a$ and $b$.
The number of the degrees of freedom of a single molecule is given by constant $c$ ($c=3/2$ for single atoms), and $k_{B}$ is the Boltzmann constant.

In equilibrium thermodynamics, $a$ and $b$ are also parameters of state just like $S$, $V$ and $N$ \citep{watanabe2004precise,watanabe2004thermodynamic,kim2016equation}.
Therefore, for the Van der Waals gas they are present in the differential of energy (first law of thermodynamics)
\begin{equation}
 dU=TdS-pdV-\frac{N^{2}}{V}da +	Nk_BT\left(\frac{V}{N}-b\right)^{-1}db
 \label{eq:diff U eq}
\end{equation}
with temperature $T=\partial U\left(S,V,a,b\right)/\partial S$,
pressure $p=-\partial U\left(S,V,a,b\right)/\partial V$, 
$\frac{N^{2}}{V} = -\partial U\left(S,V,a,b\right)/\partial a$ and $Nk_BT\left(\frac{V}{N}-b\right)^{-1} = \partial U\left(S,V,a,b\right)/\partial b$ \citep{Thermodynamics_and_an_Introduction_to_Thermostatistics_2ed_H_Callen}.
Each term in the above expression corresponds to one way the energy enters the Van der Waals gas.
$\mkern3mu\mathchar'26\mkern-12mu dQ=TdS$
is the heat, $\mkern3mu\mathchar'26\mkern-12mu dW=-pdV$ is the elementary
mechanical work when the volume changes,
and the last two terms represent the work of external sources required to change the strength
of interactions.
Modifications of an interaction parameter are used, e.g., in the thermodynamic integration methods \citep{frenkel2002understanding}.

In the following sections, we will benefit from the equivalence between the fundamental thermodynamic relation for the Van der Waals fluid (\ref{eq:fundamental eq})
and the energy differential (\ref{eq:diff U eq}) supplemented with the equations of state
\begin{subequations}
\label{eq:both eos}
\begin{align}
p &=\frac{nk_{B}T}{1-nb}-an^{2},\label{eq:pressur eos}
\\
u &= cnk_{B}T-an^{2},\label{eq:temperature eos}
\end{align}
\end{subequations}
where $n=N/V$ is particle density and $u=U/V$ is energy density.

\section{Van der Waals gas in a heat flow}

We discuss a simplified Van der Waals gas ($b=0$) first.
Consider a system schematically shown in Fig. 1, a rectangular cavity with a constant amount of particles $N$.
\begin{figure}[bp]
	\includegraphics[width=8.5cm]{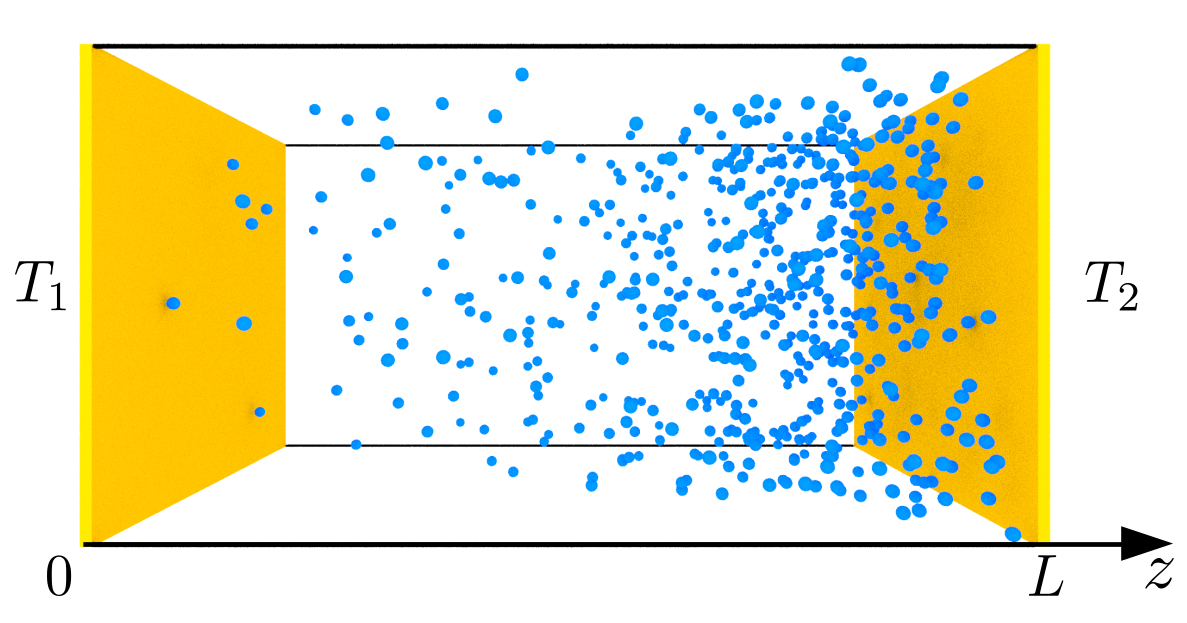}
	\caption{The schematic of the Van der Waals gas between parallel walls separated by a distance $L$. 
    The walls are kept at temperatures $T_1 > T_2$, and the density of spheres represents the variation of the gas density in the temperature gradient.}
\end{figure}
We distinguish two parallel walls separated by a distance $L$ in the $z$ direction.
The walls are kept at temperatures $T_{1}$ and $T_{2}$.
In other directions, we assume the translational invariance, which constitutes a 1D problem. 
We assume the local equilibrium,
that is, the dynamics of the gas density $n\left( z \right)$ is governed by thermo-hydrodynamic equations:
mass continuity, momentum balance and energy balance equations \citep{Groot_Mazur_Non-equilibrium_thermodynamics},
which are supplemented with equations of states (\ref{eq:both eos})
\begin{subequations}
\label{eq:nonh eos}
\begin{align}
p \left(z\right) & = n\left(z\right)k_{B}T\left(z\right)-an\left(z\right)^{2},\label{eq:nonh pressure eos}
\\
u \left(z\right)& =cn\left(z\right)k_{B}T\left(z\right)-an\left(z\right)^{2}\label{eq:nonh temperature eos}
\end{align}
\end{subequations}
valid for every coordinate $z$. 
In the steady state, inside the finite 1D segment, the velocity field has to be equal 0 everywhere.
The constant pressure solution $p\left(z\right)=\textrm{const}$ follows.
Another simplification resulting from the stationary condition is the Laplace equation for the temperature profile with linear solution
\begin{equation}
T\left(z\right)=T_{1}+\left(T_{2}-T_{1}\right)\frac{z}{L}.\label{eq:temperature profile}
\end{equation}

To determine the concentration profile,
we observe that equation (\ref{eq:nonh pressure eos}) written locally, $p=nk_{B}T-an{}^{2}$, is quadratic in density.
Thermodynamic stability conditions \citep{Thermodynamics_and_an_Introduction_to_Thermostatistics_2ed_H_Callen}
requires that $\left(\partial p/\partial n\right)_{T}\geq0$, which gives $k_{B}T-2an\geq0$.
Therefore, the only physical solution for the density that satisfies (\ref{eq:nonh pressure eos}) is given by,
\begin{equation}
n\left(z\right)=\frac{k_{B}T\left(z\right)-\sqrt{\left(k_{B}T\left(z\right)\right)^{2}-4ap}}{2a},\label{eq:density physical solution}
\end{equation}
and the stability condition, $k_{B}T\left(z\right)-2an\left(z\right)\geq0$,
with the use of the above expression for $n\left(z\right)$ is reduced to $\left(k_{B}T\left(z\right)\right)^{2}\geq4ap$.
Because the pressure in the system is constant, and the temperature profile is known, eqs.
(\ref{eq:temperature profile}) and (\ref{eq:density physical solution}) allow us to determine the total number of particles in the system,
\begin{align}
 & N\left(T_{1},T_{2},A,L,p\right)=A\int_{0}^{L}dz\,n\left(z\right)=\frac{ALk_{B}\left(T_{1}+T_{2}\right)}{2a}\times\nonumber \\
 & \times\left[\frac{1}{2}+\frac{4ap}{k_{B}^{2}\left(T_{2}^{2}-T_{1}^{2}\right)}\int_{k_{B}T_{1}/\sqrt{4ap}}^{k_{B}T_{2}/\sqrt{4ap}}du\,\sqrt{u^{2}-1}\right],
	\label{eq:N by control param}
\end{align}
where $A$ is the surface area of the system in the direction of translational invariance.
Similarly, from the eq. (\ref{eq:nonh temperature eos}) we determine the total internal energy
\begin{align}
U\left(T_{1},T_{2},A,L,p\right)=A\int_{0}^{L}dz\,u\left(z\right)\nonumber \\
=ALp\left[1+\frac{\left(c-1\right)\sqrt{4ap}}{k_{B}\left(T_{2}-T_{1}\right)}\left(g\left(\frac{k_{B}T_{2}}{\sqrt{4ap}}\right)-g\left(\frac{k_{B}T_{1}}{\sqrt{4ap}}\right)\right)\right]
\label{eq:U by control param}
\end{align}
with $g\left(x\right)=\frac{1}{3}\left[x^{3}-\left(x^{2}-1\right)^{\frac{3}{2}}-1\right]$.

\section{Net heat for Van der Walls gas and new parameter of state}

In a steady state, the same amount of heat enters through one wall and leaves through the other.
However, during the transition from one steady state to another, e.g., by a slight change of temperature $T_{2}$ or by a motion of the right wall changing $L$ (see Fig. 1), this balance is, in general, disturbed and the net heat may flow to the system changing its internal energy \citep{holyst2022thermodynamics}.
In the case of a very slow transition between stationary states,
the energy changes only by means of mechanical work and heat flow
\begin{equation}
 dU=\mkern3mu\mathchar'26\mkern-12mu dQ+\mkern3mu\mathchar'26\mkern-12mu dW.
\end{equation}
The mechanical work is given by 
\begin{equation}
\mkern3mu\mathchar'26\mkern-12mu dW=-pdV.\label{eq:work dif by pressure}
\end{equation}
and the energy balance during the transition between non-equilibrium steady states has the following form
\begin{equation}
dU=\mkern3mu\mathchar'26\mkern-12mu dQ-pdV. \label{eq:general energy balance}
\end{equation}
The above equation reduces to the first law of thermodynamics in equilibrium.
It has the same form,
but here the $\mkern3mu\mathchar'26\mkern-12mu dQ$
is the net heat transferred to the system during a small change between two stationary instead of equilibrium states.

We obtain the formal analogy between equilibrium and stationary state for the Van der Waals gas by
integrating the equations of state (\ref{eq:nonh eos}) over the volume 
\begin{subequations}
\begin{align}
pV &=A\int_{0}^{L}dz\,n\left(z\right)k_{B}T\left(z\right)-Aa\int_{0}^{L}dz\,n\left(z\right)^{2},
\\
U  &=\frac{3}{2}A\int_{0}^{L}dz\,n\left(z\right)k_{B}T\left(z\right)-Aa\int_{0}^{L}dz\,n\left(z\right)^{2},
\end{align}
\end{subequations}
and by introducing average temperature
\begin{equation}
T^{*} \equiv \frac{A\int_{0}^{L}dz\,n\left(z\right)T\left(z\right)}{A\int_{0}^{L}dz\,n\left(z\right)}  \label{eq:def Tstar}
\end{equation}
and the effective potential energy  parameter
\begin{equation}
a^{*} \equiv \frac{Aa\int_{0}^{L}dz\,n\left(z\right)^{2}}{AL\bar{n}^{2}}=\frac{a\int_{0}^{L}dz\,n\left(z\right)^{2}}{L\bar{n}^{2}}, 
\label{eq:def astar}
\end{equation}
where $\bar{n}=N/V$ is average particle density
and $\bar{u}=U/V$ is the total energy of the system divided by its volume.
As a result, we obtain two relations 
\begin{subequations}
\label{eq:eff eos}
\begin{align}
p &=\bar{n}k_{B}T^{*}-a^{*}\bar{n}^{2},
\label{eq:eff pressure eos}
\\
\bar{u} &=c\bar{n}k_{B}T^{*}-a^{*}\bar{n}^{2},
\label{eq:eff temperature eos}
\end{align}
\end{subequations}
which (for $b=0$) are formally identical to (\ref{eq:both eos}).
Because the equations (\ref{eq:eff eos}) have the same structure as the equilibrium equation
of state, they relate to the fundamental relation (\ref{eq:fundamental eq})
\begin{equation}
U\left(S^{*},V,N,a^{*}\right)=N\left(\frac{V}{N}\right)^{-\frac{1}{c}}\exp\left[\frac{S^{*}-Ns_{0}}{cNk_{B}}\right]-a^{*}\frac{N^{2}}{V}, \label{eq:nonh fundamental U via Sstar}
\end{equation}
but with effective parameters.
Moreover, the above equation defines $S^{*}$ and it has a differential
\begin{equation}
dU=T^{*}dS^{*}-pdV-\frac{N^{2}}{V}da^{*},\label{eq:diff noneq U eq}
\end{equation}
where $T^{*}=\left(\partial U/\partial S\right)_{V,N,a^{*}}$,
$p=\left(\partial U/\partial V\right)_{S^{*},N,a^{*}}$ and $\frac{N^{2}}{V} = -\partial U\left(S^{*},V,a^{*}\right)/\partial a^{*}$.

The comparison of equations (\ref{eq:diff noneq U eq}) and (\ref{eq:general energy balance})
gives the relation between the net heat in the system and the effective
entropy,
\begin{equation}
 \mkern3mu\mathchar'26\mkern-12mu dQ=T^{*}dS^{*}-\frac{N^{2}}{V}da^{*}.
\end{equation}
The net heat flow during the transition between two steady
states is a combination of the two exact differentials:
effective entropy $dS^{*}$, and effective interaction $da^{*}$.
It is contrary to equilibrium thermodynamics, where the heat is determined solely by the temperature and the change of entropy.

\section{The integrating factor for net heat in the Van der Waals gas in steady states does not exist}

We rearrange Eq. (\ref{eq:general energy balance}) to get the net heat, 
\begin{equation}
	\mkern3mu\mathchar'26\mkern-12mu dQ=dU+pdV.\label{eq:dif net heat}
\end{equation} 
The energy and pressure can be determined from the stationary solution.
Therefore we are in position to ask whether the heat differential $\mkern3mu\mathchar'26\mkern-12mu dQ$ has an integrating factor in
space $T_{1},T_{2},V$.
For the ideal gas ($a=0$) the integrating factor exists \citep{holyst2022thermodynamics}.
It follows that there exists a function of state, which is constant if
the steady state system is ``adiabatically insulated'' 
(i.e. the net heat vanishes, $\mkern3mu\mathchar'26\mkern-12mu dQ=0$).

We say that a differential form $\mkern3mu\mathchar'26\mkern-12mu dF=f_{1}\left(x_{1},x_{2},x_{3}\right)dx_{1}+f_{2}\left(x_{1},x_{2},x_{3}\right)dx_{2}+f_{3}\left(x_{1},x_{2},x_{3}\right)dx_{3}$
has an integrating factor if there exists a function $\phi\left(x_{1},x_{2},x_{3}\right)$
whose differential is related to $\mkern3mu\mathchar'26\mkern-12mu dF$
by
\[
d\phi\left(x_{1},x_{2},x_{3}\right)\equiv\mkern3mu\mathchar'26\mkern-12mu dF/\mu\left(x_{1},x_{2},x_{3}\right).
\]
The function $\mu$ is called the integrating factor and $\phi$ is
called the potential of the form $\mkern3mu\mathchar'26\mkern-12mu dF$.
The differential form may be considered in different variables, e.g.
given by $y_{i}=y_{i}\left(x_{1},x_{2},x_{3}\right)$ for $i=1,2,3$.
We will write shortly, $Y\left(X\right)$. It is straightforward to
check that when the differential form is transformed into new variables,
the integrating factor is given by, $\mu\left(X\left(Y\right)\right).$
We can choose the most convenient set of variables to find the integrating
factor of a differential form.

We considered the space of the control parameters,
$T_{1},T_{2},A,L,N$. It has been used to represent the number of
particles, $N=N\left(T_{1},T_{2},A,L,p\right)$ and the energy in
the system, $U=U\left(T_{1},T_{2},A,L,p\right)$, given by 
Eqs. (\ref{eq:N by control param}) and (\ref{eq:U by control param}).
To simplify further considerations, let's
notice that the surface area, $A$, and the length of the system,
$L$, always appear in the above relations as a product, $V=AL$.
We can reduce the space of control parameters to $T_{1},T_{2},V,N$.
Because we confined our considerations to constant number of particles,
$N$, we have three parameters, $T_{1},T_{2},V$. However, the natural
variables of the differential form (\ref{eq:dif net heat}) are $U$,
$V$. We will use them in the following considerations and we take
$\tau=T_{2}/T_{1}$ as the third parameter.

Suppose that the net heat has the integrating factor. It means that
there exists a potential $\phi\left(U,V,\tau\right)$ which differential
is related to the net heat differential by
\[
d\phi\left(U,V,\tau\right)\equiv\mkern3mu\mathchar'26\mkern-12mu dQ/\mu\left(U,V,\tau\right).
\]
By definition, $d\phi=\frac{\partial\phi}{\partial U}dU+\frac{\partial\phi}{\partial V}dV+\frac{\partial\phi}{\partial\tau}d\tau$.
On the other hand the above relation with Eq. (\ref{eq:dif net heat})
gives, $d\phi=1/\mu\left(U,V,\tau\right)dU+p\left(U,V,\tau\right)/\mu\left(U,V,\tau\right)dV.$
Equality of the second derivatives for all three independent variables
$U,V,\tau$ is a necessary condition for the existence of $\phi$.
It is easy to check that this condition is satisfied only if $p\left(U,V,\tau\right)$
does not depend on $\tau$,
\[
\left(\frac{\partial p}{\partial\tau}\right)_{U,V}=0.
\]
Equivalently, if $\left(\partial p/\partial\tau\right)_{U,V}\neq0$,
then the integrating factor of the net heat does not exist.

The above condition requires the determination of $p\left(U,V,\tau\right)$.
The pressure can be determined from 
Eqs. (\ref{eq:N by control param}) and (\ref{eq:U by control param}), 
which have the following form, $N=N\left(T_{1},T_{2},V,p\right)$,
and, $U=U\left(T_{1},T_{2},V,p\right)$. Inversion of the former relation
would lead to the formula $p=p\left(T_{1},T_{2},V,N\right)$, but
we are not able to obtain explicit expression for $p$ in terms of
elementary functions. However, what we need is not the function itself,
but its derivative over $\tau$. Even if a function is given implicitly,
its derivative can be explicitly determined with the use of the simple
properties of derivatives \citep{Thermodynamics_and_an_Introduction_to_Thermostatistics_2ed_H_Callen}.
We have a similar situation here: although $p\left(U,V,\tau,N\right)$
with $\tau=T_{2}/T_{1}$ cannot be explicitly determined from $N=N\left(T_{1},T_{2},V,p\right)$,
and, $U=U\left(T_{1},T_{2},V,p\right)$, but its derivative, $\left(\partial p/\partial\tau\right)_{U,V}\neq0$,
can be determined explicitly. By using properties of derivatives of
functions $U=U\left(T_{1},T_{2},V,p\right)$ and $N=N\left(T_{1},T_{2},V,p\right)$
one shows the following property. The derivative $\left(\partial p/\partial\tau\right)_{U,V}\neq0$
does not vanishes, if the following conditions are satisfied:
\begin{equation}
	\left\{ U,N\right\} _{T_{1},T_{2}} \neq 0 \label{eq:poisson}
\end{equation}
and
\[
\frac{T_{2}}{T_{1}}\left\{ U,N\right\} _{p,T_{2}}+\left\{ U,N\right\} _{p,T_{1}}\neq0.
\]
In the above expressions the Poisson bracket is defined by $\left\{ f,g\right\} _{x,y}\equiv\partial f/\partial x\,\partial g/\partial y-\partial g/\partial x\,\partial f/\partial y$.
The proof of the above property requires standard properties of derivatives
under change of variables \citep{Thermodynamics_and_an_Introduction_to_Thermostatistics_2ed_H_Callen}
and is omitted here.

It can be directly checked whether the Poisson bracket (\ref{eq:poisson})
does not vanish for functions $U=U\left(T_{1},T_{2},V,p\right)$ and
$N=N\left(T_{1},T_{2},V,p\right)$ given by Eqs. (\ref{eq:N by control param}) and (\ref{eq:U by control param}).
Calculations are straightforward but cumbersome. To convince
the reader that the Poisson bracket (\ref{eq:poisson}) does not vanish,
we consider the limit $T_{2}\to T_{1}.$ It gives the following expression,
\begin{align}
\lim_{T_{2}\to T_{1}}\frac{\partial}{\partial T_{2}}\left\{ U,N\right\}_{T_{1},T_{2}}  = \nonumber \\
=\frac{(c-1)k_{B}^{3}V^{2}\left(\frac{k_{B}T_{1}}{\sqrt{ap}}-\sqrt{\frac{\left(k_{B}T_{1}\right)^{2}}{ap}-4}\right)}{8a^{2}\left(\frac{\left(k_{B}T_{1}\right)^{2}}{ap}-4\right)^{3/2}}.
\end{align}
It follows that even in the neighborhood of the equilibrium state,
$T_{2}\approx T_{1}$, the above Poisson bracket does not vanish.
As a consequence, the heat differential for Van der Waals gas has
no integrating factor. Thus a function that plays the role of entropy does not exist for
Van der Waals gas in a steady state with heat flow. The representation
$\mkern3mu\mathchar'26\mkern-12mu dQ=T^{*}dS^{*}$ is impossible.

\section{Global steady thermodynamics for Van der Walls gas with $b\protect\neq0$}

So far we have introduced global steady thermodynamic description
for Van der Walls gas given by Eq. (1) with reduced parameter, $b=0$.
Here we consider $b\neq0$ case in which the following equations of
state

\begin{equation}
	p=\frac{n\left(z\right)k_{B}T\left(z\right)}{1-bn\left(z\right)}-an\left(z\right)^{2},\label{eq:b nonh pressure}
\end{equation}
\begin{equation}
	u\left(z\right)=cn\left(z\right)k_{B}T\left(z\right)-an\left(z\right)^{2},\label{eq:b nonh temperature}
\end{equation}
describe Van der Walls gas in a stationary state. As before, the pressure
in the system is constant. Integration of the above equations over
volume leads to the following relations,
\begin{equation}
	p=\frac{\bar{n}k_{B}T^{*}}{1-\bar{n}b^{*}}-a^{*}\bar{n}^{2},\label{eq:eff pressure eos-2}
\end{equation}
\begin{equation}
	\bar{u}=c\bar{n}k_{B}T^{*}-a^{*}\bar{n}^{2},\label{eq:eff temperature eos-2}
\end{equation}
where $T^{*}$ and $a^{*}$ are defined by Eqs. (\ref{eq:def Tstar}) and (\ref{eq:def astar}) while
$b^{*}$ is defined by the following formula
\begin{equation}
	\frac{\bar{n}k_{B}T^{*}}{1-\bar{n}b^{*}}=\frac{1}{L}\int_{0}^{L}dz\frac{n\left(z\right)k_{B}T\left(z\right)}{1-bn\left(z\right)}.\label{eq:def b star}
\end{equation}

Eqs. (\ref{eq:eff pressure eos-2}) and (\ref{eq:eff temperature eos-2})
show that the nonhomogeneous Van der Waals gas in a stationary state
with a heat flow can be mapped on the homogeneous Van der Waals gas
with effective temperature and interaction parameters, $T^{*},a^{*},b^{*}$.
Therefore it has the following fundamental relation (1),
\begin{equation}
	U=N\left(\frac{V}{N}-b^{*}\right)^{-\frac{1}{c}}\exp\left[\frac{S^{*}-Ns_{0}}{cNk_{B}}\right]-a^{*}\frac{N^{2}}{V},\label{eq:effective fundamental eq}
\end{equation}
with partial derivatives, $T^{*}=\partial U\left(S^{*},V,a^{*},b^{*}\right)/\partial S^{*}$
and $p=-\partial U\left(S^{*},V,a^{*},b^{*}\right)/\partial V$. Differential
of the above fundamental equation gives,

\begin{equation}
	dU=T^{*}dS^{*}-pdV-\frac{N^{2}}{V}da^{*}+
	Nk_BT^*\left(\frac{V}{N}-b^{*}\right)^{-1}db^{*} .\label{eq:diff U eq-2}
\end{equation}
Using also the expression for the net heat (\ref{eq:dif net heat}), we identify the heat
differential,
\[
\mkern3mu\mathchar'26\mkern-12mu dQ=T^{*}dS^{*}-\frac{N^{2}}{V}da^{*}+Nk_BT^*\left(\frac{V}{N}-b^{*}\right)^{-1}db^{*}.
\]
The above equations describe the energy balance for Van der Walls
gas with a heat flow and they correspond to the first law in equilibrium
thermodynamics when the heat flow vanishes.

The parameters $T^{*},a^{*},b^{*}$ defined by Eqs. (\ref{eq:def Tstar}-\ref{eq:def b star})
are not independent. To explain it, we keep in mind that for a given
number of particles, three control parameters $T_{1},T_{2},V$ are
sufficient to determine the system's energy, work, and net heat differential.
On the other hand, the energy differential in Eq. (\ref{eq:diff U eq-2})
is given by four parameters, $S^{*},V,a^{*},b^{*}$. It follows that
$S^{*},V,a^{*},b^{*}$ are dependent. Consequently, one of these parameters
should be determined by the others, e.g. $b^{*}=b^{*}\left(S^{*},V,a^{*}\right)$.

In the above considerations, Van der Waals gas was enclosed between two parallel walls. Control parameters
$T_1$, $T_2$, $V$, and $N$ determine the steady state. In a more practical situation, the system does not need to be rectangular,
and several temperature parameters, $T_1,\ldots,T_N $, determine the boundary conditions.
The same procedure determines the fundamental relation (\ref{eq:effective fundamental eq}) because 
it applies to any density and temperature profile. 
Even in a situation with an arbitrary number of control parameters ($N>2$), 
the five parameters of states $S^*$, $V$, $N$, $a^*$ and $b^*$ are sufficient to determine the energy exchange in the system.

\section{Summary}

A fundamental relation such as Eq. (\ref{eq:fundamental eq})
plays a key role in equilibrium thermodynamics. The fundamental relation, by definition, is a relation
between parameters of the system's state, from which one can ascertain
all relevant thermodynamic information about the system \citep{Thermodynamics_and_an_Introduction_to_Thermostatistics_2ed_H_Callen}.
It includes the identification of different forms of energy exchange of the system with the environment.
In equilibrium thermodynamics the particular terms of the energy differential
correspond to heat, mechanical work, or chemical work.
In the same spirit, Eq. (\ref{eq:effective fundamental eq}) is the
fundamental relation for the Van der Waals gas in a steady state with
a heat flow. Its differential (\ref{eq:diff U eq-2}) gives information
about the net heat and the work performed on the system. Eq. (\ref{eq:diff U eq-2})
directly reduces to the first law of thermodynamics when the heat
flow vanishes. It represents the first
law of the global steady thermodynamic description of an interacting
system subjected to heat flow.

The integrating factor for the heat differential in
the case of the ideal gas discussed previously \citep{holyst2022thermodynamics}
allowed us to introduce the non-equilibrium entropy and use it to
construct the minimum energy principle beyond equilibrium. This principle
generalizes thermodynamics' second law beyond equilibrium. Here we
showed that the net heat has no integrating factor. It excludes a
direct generalization of the second law along the line proposed in
\citep{holyst2022thermodynamics}. However, it does not exclude a
possibility that such a principle also exists in the case of an interacting
gas.

This paper suggests a general prescription for formulating
the fundamental relation of global nonequilibrium steady thermodynamics.
First, we identify equilibrium equations of state. Next, we write the local equations of state.
Whether these equations are in the same form in equilibrium thermodynamics or some other form remains to be found.
Next, we average these local (or non-local) equations of the state over the entire system.
We insist that the global equations of a nonequilibrium state should have the same form as at equilibrium but with new state parameters.
These parameters emerge after averaging the local equations over the entire system.
In the case of Van der Waals, the new state parameters emerged, $a^*$ and $b^*$.
These parameters are constant at equilibrium since they are material parameters that define interactions in a particular system.
This result suggests that, in general, all material parameters in the equilibrium equations of states will become parameters of state in the nonequilibrium systems.

\section*{Acknowledgements}
P. J. Z. would like to acknowledge the support of a project that has received funding from the European Union`s Horizon 2020 research and innovation
program under the Marie Sklodowska-Curie grant agreement No. 847413 and was a part of 
an international co-financed project founded from the program of the 
Minister of Science and Higher Education entitled `PMW' in the years 2020-2024; agreement No. 5005/H2020-MSCA-COFUND/2019/2.

\bibliographystyle{unsrt}

\end{document}